\documentclass[aps,pre,preprint,showpacs]{revtex4}
\usepackage{amsmath}
\usepackage[pdftex]{hyperref}
\usepackage{graphicx}% include figure files
\usepackage{subfigure}
\usepackage{epsf}
\usepackage{dcolumn}% align table columns on decimal point
\usepackage{bm}% bold math
\usepackage{color}% color

\begin{document}

\title{Continuous Time Random Walk with time-dependent jump probability : A Direct Probabilistic Approach}

\author{Shovan Dutta,$^1$ Subhankar Ray,$^2$ and J. Shamanna$^3$}

\affiliation{$^1$ Department of Electronics and Telecommunication Engineering, Jadavpur University, Calcutta 700 032, India}
\affiliation{$^2$ Department of Physics, Jadavpur University, 
Calcutta 700 032, India}
\email{sray@phys.jdvu.ac.in}
\affiliation{$^3$ Physics Department, University of Calcutta, Calcutta 700 009, India}
\email{jlsphy@caluniv.ac.in}

\begin{abstract}
We investigate the dynamics of a particle executing a general Continuous Time Random Walk (CTRW) in three dimensions under the influence of arbitrary time-varying external fields.  Contrary to the general approach in recent works, our method invokes neither the Fractional Fokker-Planck equation (FFPE) nor the Stochastic Langevin Equation (SLE). Rather, we use rigorous probability arguments to derive the general expression for moments of all orders of the position probability density of the random walker for arbitrary waiting time density and jump probability density. Closed form expression for the position probability density is derived for the memoryless condition. For the special case of CTRW on a one-dimensional lattice with nearest neighbour jumps, our equations confirm the phenomena of ``death of linear response" and ``field-induced dispersion" for sub-diffusion pointed out in [I. M. Sokolov and J. Klafter, Phys. Rev. Lett. {\bf 97}, 140602 (2006)]. However, our analysis produces additional terms in the expressions for higher moments, which have non-trivial consequences. We show that the disappearance of these terms result from the approximation involved in taking the continuum limit to derive the generalized Fokker-Planck equation. This establishes the incompleteness of the FFPE formulation, especially in predicting the higher moments. We also discuss how different predictions of the model alter if we allow jumps beyond nearest neighbours and possible circumstances where this becomes relevant.

\end{abstract}

\maketitle

\section{Introduction}
In the past few decades, an increasing number of natural phenomena in diverse branches of physics, biology and the related fields of science have been shown to exhibit anomalous diffusion. Examples of processes displaying subdiffusive dynamics include gel electrophoresis of polymers (e.g., DNA), transport in amorphous semiconductors \cite{ScherMontroll, ScherLax}, nuclear magnetic resonance, motion of proteins through cell membranes, diffusion in percolative and porous systems, transport in fractal geometries, contaminants transport by underground water and many others (see \cite{review} for interesting reviews). The Continuous Time Random Walk (CTRW) model introduced by Montroll and Weiss \cite{MW, Weiss, Hughes}, with power-law waiting time distributions lacking first moment, provides a powerful mathematical tool to describe sub-diffusion \cite{ScherMontroll, shlesinger}. The non-stationary nature of these (non-Markovian) subdiffusive CTRWs leads to the effects of aging in various systems \cite{aging, physica, adhoc}. In the absence of time-dependent external fields, the CTRW process can be subordinated to the simple random walk, which enables one to describe it within the framework of Fractional Fokker-Planck Equations (FFPEs) \cite{review, FFPE, subordination, linres, pp}. The FFPE can also be derived from the analysis of a (ordinary) Langevin equation in subordinated random operational time \cite{langevin1, langevin2}.

%Many physical systems exhibit a wide array of interesting features when subjected to an external potential varying in time. Therefore, it is tempting to investigate subdiffusive phenomena in time-dependent force fields.%
However, the generalization of the above scheme to situations where an external time-varying field is present is non-trivial because the force physically changes in real time which cannot be transformed to the random operational time, thus creating difficulties in subordination \cite{goychuk, heinsalu1}. It has been shown that the the FFPE for the time-independent case, when generalized {\it ad hoc} to a time-dependent force \cite{physica, adhoc}, does not correspond to physical reality \cite{heinsalu1}. Modified FFPEs for time-dependent forces were derived in Refs. \cite{PRL, heinsalu1} but later more rigorous treatments revealed that they were valid only for some particular deterministic time-dependences \cite{PRE,heinsalu2}. Recently, a subordinated stochastic Langevin equation \cite{spacetime1} and an FFPE \cite{spacetime2} have been proposed for modelling sub-diffusion in space-and-time-dependent force fields.

In spite of these advances, the theoretical investigation of sub-diffusion in a time-dependent force field still presents a challenge that is far from trivial. Some of the difficulties and ambiguities were pointed out in Ref. \cite{heinsalu2}. First, it is known that there is no unique physical mechanism behind the occurrence of sub-diffusion in condensed media \cite{nonunique}. Second, in the CTRW model, successive waiting times are assumed to be mutually uncorrelated, which may not be true in practice \cite{correlation}. Third, as the time-varying field will alter the rates for moving in different directions when escaping from a trap location, the waiting time distribution will generally be affected by the field \cite{RTD2}. However, due to our lack of knowledge of a concrete mechanism, it is not clear to what extent this influence will change the predictions based on a decoupled model where the waiting time distribution is assumed to be independent of the external perturbation \cite{RTD1}.

The limitation we want to stress in this paper is that even for an uncorrelated decoupled CTRW model, the FFPE formulation is incomplete \cite{heinsalu2}, as no (single event) non-Markovian master equation can fully characterize the underlying (multitime event) non-Markovian stochastic process which lacks the factorization property \cite{incomplete1}. Therefore, neither a non-Markovian FFPE nor its solution can display all the features of the non-Markovian CTRW \cite{incomplete2}. Further, the general strategy to form an FFPE adopted in all relevant works (see, for example \cite{PRL}) is to first write down the generalized master equation (GME) for a CTRW on a one-dimensional lattice with jumps restricted to nearest neighbours, and then take its continuum limit. This method is not rigorous and as we show in this article, the predictions of the FFPE regarding the higher moments of the position probability density of the random walker increasingly differ from those of the GME. Also, if the CTRW is not confined in a single dimension, there are several non-unique ways of taking the continuum limit corresponding to different lattices.

In this paper, we treat the problem of purely time-dependent external fields. Contrary to the customary approach, we start with a general CTRW in a three-dimensional continuum and derive expressions for the moments of all orders of the position probability density $p(\vec{r},t)$ of the walker for arbitrary waiting time density $\psi (t)$ and jump probability density $\phi (\vec{r}|t)$. Later on, we choose a particular form of $\phi (\vec{r}|t)$ to recover the special case of the one-dimensional lattice. We find that in this case, our equations do predict the same mean position and standard deviation as those in Refs. \cite{PRL, heinsalu1, heinsalu2, PRE}, thus confirming the phenomena of ``field-induced dispersion'' and ``death of linear response'' for systems showing sub-diffusion. However, for the higher moments, our analysis produces extra terms in addition to those in the aforesaid references, which have non-trivial consequences. We point out that the disappearance of these additional terms is an artifact of the approximation involved in taking the continuum limit to derive the generalized FFPE. These terms are recovered if one calculates moments directly from the GME. Our method is based on direct probability arguments \cite{Feller} and some well-known properties of Integral transforms and does not invoke the fractional Fokker-Planck equation. This is a key departure from other recent works on the subject. It presents a rigorous treatment of the general CTRW in three-dimensions. Consequently, the results are valid for any time-dependence of the external field within the assumptions of uncorrelated decoupled CTRW. We find the interesting result that the $n$th moment of $p(\vec{r},t)$ depends on the $m$th moment of $\phi (\vec{r} | t)$ if and only if $m \leq n$. For sub-diffusive waiting time densities,  $\psi(t) \sim t^{-1-\alpha}$ with $0 < \alpha < 1$, we show how the normal approximation breaks down more and more as one lowers $\alpha$ towards zero. We describe a one-parameter family of probability density functions which can well-approximate the position probability density of the random walker for large values of $t$. For the special case where the random walker is memoryless, i.e., where the probability of making a jump does not depend on the time at which the previous jump was made, the process becomes Markovian. In this instance, we are able to derive $p(\vec{r},t)$ in closed form and show how its moments are simply related to the moments of $\phi (\vec{r}|t)$. We conclude with a discussion on the significance of choices of $\phi (\vec{r}|t)$ for which the jumps are not restricted to nearest neighbours and a brief note on some open problems in the subject.

\section{Formulation of the Problem}
Let us first describe the model of the three-dimensional random walk adopted in this paper. The random walker, on arriving at a position $\vec{\rho}$ at time $\tau$ stays there for a sojourn time $t^{'}$, after which it jumps to a position $\vec{r}$ at time $t=\tau + t^{'}$. The waiting time $t^{'}$ is distributed with the probability density function $\psi (t^{'})$. The jumps are assumed to be instantaneous on the time scale of typical waiting times and changes of external parameters. A physical justification of this assumption may be found in the case of sub-diffusion, where long waiting times are often dominant. The probability that a jump occurring at time $t$ results in a displacement between $\vec{r^{'}}$ and $\vec{r^{'}} + d \vec{r^{'}}$ is given by $\phi (\vec{r^{'}}| t) d^3 \vec{r^{'}}$. We seek to find the probability density $p (\vec{r} , t)$ that the random walker is found at position $\vec{r}$ at time $t$. Also, we choose our references in  such a manner that at $t=0$, the walker has just arrived at $\vec{r}=\vec{0}$.

Let us define $\Psi (t)$ as the probability that the time interval between two successive jumps is greater than $t$. Clearly,
\begin{equation}
\Psi (t) = \int_{t}^{\infty} \psi (t^{'}) d t^{'}\; .
\label{Psi}
\end{equation}

We shall require another intermediate probability density $f(\vec{r},t)$ defined by,
\begin{align}
\nonumber f(\vec{r},t) d^3 \vec{r} d t  =& \;\mbox{probability that the walker arrives between $\vec{r}$ and $\vec{r} + d \vec{r}$} \\
\nonumber & \;\mbox{within the time interval $t$ to $t + d t$ after one or more jumps.}
\label{f}
\end{align}

Then the desired probability density $p(\vec{r},t)$ can be written as a sum of two parts,
\begin{eqnarray}
p(\vec{r},t) &=& \delta(\vec{r}) \Psi(t) + q(\vec{r},t) \; ,\label{divide} \\
\mbox{where,} \;\;\; q(\vec{r},t) &=& \int_{0}^{t} f(\vec{r},\tau) \Psi(t - \tau) d \tau \; .\label{conv1}
\end{eqnarray}

The first term involving the delta function expresses the possibility that no jump has taken place since $t=0$, in which case, the random walker is bound to remain at $\vec{r}=\vec{0}$. Conversely, the probability density of being at $\vec{r}$ at time $t$ after executing at least one jump since $t=0$ is given by $q(\vec{r},t)$, which can be written as a convolution integral as in (\ref{conv1}). Here $\tau$ denotes the time of occurrence of the last jump which took the walker to $\vec{r}$ and the second factor $\Psi(t - \tau)$ ensures that no jump has taken place since then. From (\ref{divide}) and (\ref{conv1}), it is evident that our problem reduces to the determination of the intermediate density $f(\vec{r},t)$.

This intermediate density can again be broken into two parts. It is possible that the only jump took place at time $t$, taking the walker to $\vec{r}$, for which the probability density is given by $\psi(t) \phi(\vec{r}|t)$. Alternatively, the walker reached some position $\vec{\rho}$ at some time $\tau$ lying between $0$ and $t$, and was transferred to $\vec{r}$ in the last jump at time $t$. This leads us to the following integral equation for $f(\vec{r},t)$.
\begin{equation}
f(\vec{r},t) = \psi(t) \phi(\vec{r}|t) + \int_{-\infty}^{\infty} \int_{0}^{t} f(\vec{\rho},\tau) \psi(t-\tau) \phi(\vec{r}-\vec{\rho} | t) d \tau d^3 \rho \; .
\label{recursion}
\end{equation}

This equation can be simplified by taking the Fourier transform with respect to $\vec{r}$ and using the convolution theorem,
\begin{equation}
\bar{f}(\vec{k},t) = \bar{\phi}(\vec{k}|t) \psi(t) + \bar{\phi}(\vec{k}|t) \int_{0}^{t} \bar{f}(\vec{k},\tau) \psi(t-\tau) d \tau \; .
\label{recursion_k}
\end{equation}

For the special case where $\phi(\vec{r}|t)$ is time-independent, it is possible to solve this equation simply by taking the Laplace transform with respect to $t$, which then leads directly to the Montroll-Weiss equation \cite{MW}. However, for a non-trivial time-dependence in $\phi(\vec{r}|t)$, no such straightforward method seems to exist. In the following, we first assume a specific form of the waiting time density. After discussing its implications, we proceed to the general case.

\section{Results for Memoryless Random Walk}
In a memoryless random walk, the probability of a jump in any infinitesimal time interval $dt$ is a constant and given by $dt / T$, where $T$ is the average time interval between two successive jumps. Then the probability that no jump occurs in an interval of duration $t$ is given by,
\begin{equation}
\Psi(t) = \lim_{\delta t \rightarrow 0} \Big(1 - \frac{\delta t}{T}\Big)^{t / \delta t} = \;e^{-t / T} \; .
\end{equation} 

Using this result in (\ref{Psi}), we obtain the waiting time density $\psi(t)$,
\begin{equation}
\psi(t) = -\frac{d}{dt} \Psi(t) = \frac{1}{T} e^{-t/T}\; .
\end{equation}

Substituting the above in (\ref{conv1}) and (\ref{recursion_k}), we get,
\begin{eqnarray}
\bar{q}(\vec{k},t) &=& e^{-t/T} \int_{0}^{t} \bar{f}(\vec{k},\tau) e^{\tau /T} d \tau \label{q_ml} \; ,\\
\bar{f}(\vec{k},t) &=& \frac{1}{T} \bar{\phi}(\vec{k}|t) e^{-t/T} \Bigg[1 +  \int_{0}^{t} \bar{f}(\vec{k},\tau) e^{\tau/T} d \tau \Bigg]\; .
\label{f_ml}
\end{eqnarray}

From these two equations, one can readily derive the following,
\begin{equation}
\bar{q}(\vec{k},t) = \frac{1}{T} e^{-t/T} \int_{0}^{t} \bar{\phi}(\vec{k}|\tau) \big[1 + \bar{q}(\vec{k},\tau) e^{\tau/T}\big] d \tau \; .
\label{find_q}
\end{equation}

Differentiation with respect to $t$ yields,
\begin{equation}
\frac{\partial}{\partial t} \bar{q}(\vec{k},t) + \frac{1}{T} (1 - \bar{\phi}(\vec{k}|t))\; \bar{q}(\vec{k},t) = \frac{1}{T} e^{-t/T} \bar{\phi}(\vec{k}|t) \; .\label{IF}
\end{equation}

This can easily be solved using standard methods. After substituting the boundary condition $\bar{q}(\vec{k},0) = 0$, which follows from (\ref{q_ml}), we obtain,
\begin{equation}
\bar{q}(\vec{k},t) = e^{-t/T} \Big[\exp \Big(\frac{1}{T} \int_{0}^{t} \bar{\phi}(\vec{k}|\tau) d \tau \Big) - 1 \Big]\; .
\label{IF_sol}
\end{equation}

To obtain $\bar{p}(\vec{k},t)$, we take the Fourier transform of (\ref{divide}) and substitute the above expression for $\bar{q}(\vec{k},t)$. This gives,
\begin{equation}
\bar{p}(\vec{k},t) = e^{-t/T} \exp\Big(\frac{1}{T} \int_{0}^{t} \bar{\phi}(\vec{k}|\tau) d \tau \Big) \; .\label{p_ml}
\end{equation}

This formula can be verified in some simple cases. First, for very small values of $t$, this can be expanded as,
\begin{align}
\nonumber \bar{p}(\vec{k},t) =& \Big(1 - \frac{t}{T}\Big)\Big(1 + \frac{t}{T} \bar{\phi}(\vec{k}|0)\Big) + \mathcal{O}(t^2) \\
\nonumber =& \Big(1 - \frac{t}{T}\Big) . 1 + \frac{t}{T} \bar{\phi}(\vec{k}|0) + \mathcal{O}(t^2) \; .
\end{align}

Taking the inverse Fourier transform, one gets,
\begin{equation}
p(\vec{r},t) = \underbrace{\Big(1 - \frac{t}{T}\Big) \delta(\vec{r})}_\text{no jump} + \underbrace{\frac{t}{T} \bar{\phi}(\vec{r}|0)}_\text{one jump} + \mathcal{O}(t^2) \; .
\end{equation}

A little reflection reveals that the first (or second) term on the right hand side is nothing but the contribution of the possibility that no jump (or a single jump) occurs in time $t$.

Consider another simple case where every jump results in a constant displacement $\vec{a}$ of the walker, i.e., $\phi(\vec{r}|t)=\delta(\vec{r} - \vec{a})$. Then $\bar{\phi}(\vec{k}|t) = e^{-i\vec{k}.\vec{r}}$ and (\ref{p_ml}) gives, after expansion of the exponential,
\begin{eqnarray}
\nonumber \bar{p}(\vec{k},t) &=& \sum_{n=0}^\infty \frac{t^n}{n!} \frac{e^{-t/T}}{T^n} e^{-i\vec{k}.n\vec{a}}\; . \\
\mbox{Therefore,}\;\;  p(\vec{r},t) &=& \sum_{n=0}^\infty \frac{t^n}{n!} \frac{e^{-t/T}}{T^n} \delta(\vec{r} - n\vec{a}) \; .
\label{constant_dis}
\end{eqnarray}

Now, the probability that exactly $n$ jumps take place in duration $t$ is given by,
\begin{align}
\mathcal{P}_n(t) =& \int_{0}^{t} \int_{0}^{\tau_n}\dots \int_{0}^{\tau_3} \int_{0}^{\tau_2} \psi(\tau_1) \psi(\tau_2 - \tau_1) \dots \psi(\tau_n - \tau_{n-1}) \Psi(t - \tau_n) d \tau_1 d \tau_2 \dots d \tau_{n-1} d \tau_n \label{njumps}\\
\nonumber =& \;\frac{e^{-t/T}}{T^n} \int_{0}^{t} \int_{0}^{\tau_n}\dots \int_{0}^{\tau_3} \int_{0}^{\tau_2} d \tau_1 d \tau_2 \dots d \tau_{n-1} d \tau_n \\
\nonumber =&  \;\frac{t^n}{n!} \frac{e^{-t/T}}{T^n}\; .
\end{align}

Comparing this with (\ref{constant_dis}), we see that $p(\vec{r},t)$ can be written as,
\begin{equation}
\nonumber p(\vec{r},t) = \sum_{n=0}^{\infty} \mathcal{P}_n(t) \delta(\vec{r} - n\vec{a})\;,
\end{equation}

exactly what it should be. 

Thus, for the memoryless random walk, given the jump probability density $\phi(\vec{r}|t)$, (\ref{p_ml}) can be employed to calculate $\bar{p}(\vec{k},t)$. Then, in principle, $p(\vec{r},t)$ can be obtained via the inverse Fourier transform. However, because $\bar{\phi}(\vec{k}|t)$ is present as an exponent, even for simple choices of $\phi(\vec{r}|t)$, the inverse Fourier transform may not be available in closed form. Nevertheless, moments of all orders of the distribution $p(\vec{r},t)$ can be calculated from $\bar{p}(\vec{k},t)$ itself and these moments are what we can measure experimentally.

From the expression of the Fourier transform
\begin{equation}
\nonumber \bar{p}(\vec{k},t) = \int_{-\infty}^{\infty} p(\vec{r},t) e^{-i\vec{k}.\vec{r}} d^3 \vec{r}\; ,
\end{equation}

one can write the $n$th moment of $p(\vec{r},t)$ as,
\begin{equation}
<x^n> \;= \int_{-\infty}^{\infty} x^n p(\vec{r},t) d x = i^n \frac{\partial^n}{\partial k_x^n} \bar{p}(\vec{k},t) \Big|_{\vec{k}=\vec{0}} \; .
\label{moments}
\end{equation}

Inserting the expression of $\bar{p}(\vec{k},t)$ for memoryless random walk (see (\ref{p_ml})), one finds,
\begin{eqnarray}
<x^n>\;&=&\;i^n e^{-t/T} \frac{\partial^n}{\partial k_x^n} e^{g(\vec{k},t)} \Big|_{\vec{k}=\vec{0}} \label{moment_ml} \\
\mbox{where,}\;\; g(\vec{k},t) &=& \frac{1}{T} \int_{0}^{t} \bar{\phi}(\vec{k}|\tau) d \tau \; .\label{g}
\end{eqnarray}

The $n$th derivative in (\ref{moment_ml}) can be written as a sum of terms involving $m$th derivatives of $g(\vec{k},t)$ with $m \leq n$ (assuming both $m$ and $n$ to be nonnegative integers), as in the following,
\begin{align}
\nonumber n=1\;:&\;e^g g' \\
\nonumber n=2\;:&\;e^g(g'^2 + g'') \\
n=3\;:&\;e^g(g'^3 + 3 g' g'' + g''') \label{derivatives}\\
\nonumber n=4\;:&\;e^g(g'^4 + 6 g'^2 g'' + 4 g' g''' + 3 g''^2 + g'''') \\
\nonumber \vdots \;\; \mbox{etc. .}
\end{align}

In these expressions, $g'$ denotes the derivative of $g(\vec{k},t)$ with respect to $k_x$. Adopting similar conventions, we can write, from (\ref{g}),
\begin{align}
\nonumber g^{(n)} \big|_{\vec{k}=\vec{0}}  =& \;\frac{1}{T} \int_{0}^{t} \bar{\phi}^{(n)}(\vec{k}|\tau) \big|_{\vec{k}=\vec{0}} d \tau \\
=& \;\frac{1}{i^n} \frac{1}{T} \int_{0}^{t} <x^n>_{\phi} d \tau \qquad \mbox{where,}\;\; <x^n>_{\phi}\; = \int_{-\infty}^{\infty} x^n \phi(\vec{r}|t) d x \; .\label{xnphi}
\end{align}

In deriving the second equality, we have made use of the relation expressed by (\ref{moments}). Also,
\begin{align}
\nonumber g(\vec{0},t) =& \frac{1}{T} \int_{0}^{t} \bar{\phi}(\vec{0}|\tau) d \tau \\
=& \frac{1}{T} \int_{0}^{t} \int_{-\infty}^{\infty} \phi(\vec{r}|t) d^3 \vec{r} d \tau = \frac{t}{T}\; .
\label{g0}
\end{align}

The last equality follows from the fact that $\phi(\vec{r}|t)$ denotes a probability density and hence $\int_{-\infty}^{\infty} \phi(\vec{r}|t) d^3 \vec{r}$ must be equal to $1$. Equations (\ref{moment_ml}) to (\ref{g0}) enable us to write down the following compact relations which illustrates very clearly how the moments of $p(\vec{r},t)$ depend on those of the jump probability density.
\begin{eqnarray}
\nonumber <x> \;&=&\; \widetilde{<x>_{\phi}} \\
\nonumber <x^2>\;&=&\;\widetilde{<x>_{\phi}}^2 + \widetilde{<x^2>_{\phi}} \qquad\qquad \mbox{Thus, standard deviation}\;\;\Delta x = \sqrt{\widetilde{<x^2>_{\phi}}} \;.\\
<x^3> \;&=&\; \widetilde{<x>_{\phi}}^3 + 3 \widetilde{<x>_{\phi}} \widetilde{<x^2>_{\phi}} + \widetilde{<x^3>_{\phi}} \label{compact}\\
\nonumber <x^4> \;&=&\; \widetilde{<x>_{\phi}}^4 + 6 \widetilde{<x>_{\phi}}^2 \widetilde{<x^2>_{\phi}} + 4 \widetilde{<x>_{\phi}} \widetilde{<x^3>_{\phi}} + 3 \widetilde{<x^2>_{\phi}}^2 + \widetilde{<x^4>_{\phi}} \\
\nonumber & \vdots & \;\mbox{and so on. Here, $\widetilde{a} \equiv \frac{1}{T} \int_{0}^{t} a(\tau) d \tau$.}
\end{eqnarray}

Similar equations exist for $y$ and $z$ as well. These lead us to the following expressions for the mean position and standard deviation in three dimensions.
\begin{eqnarray}
<\vec{r}> \;&=&\; <x>\hat{i} + <y>\hat{j} + <z>\hat{k} \;=\; \widetilde{<\vec{r}>_{\phi}}\; . \\
\nonumber \Delta r \;&=&\; \sqrt{<r^2> - |<\vec{r}>|^2} \\
\nonumber                 &=&\; \sqrt{<x^2> + <y^2> + <z^2> - |\widetilde{<\vec{r}>_{\phi}}|^2} \\
                                   &=&\; \sqrt{\widetilde{<r^2>_{\phi}}}\; .
\end{eqnarray}

The relations in (\ref{compact}) can be more meaningfully expressed in terms of the central moments $\mu_n$ defined as, $\mu_n = <(x - <x>)^n>$, as follows,
\begin{equation}
\mu_1 = \widetilde{<x>_{\phi}}\;,\;\;\mu_2 = \widetilde{<x^2>_{\phi}}\;,\;\;\mu_3 = \widetilde{<x^3>_{\phi}}\;,\;\;\mu_4 = \widetilde{<x^4>_{\phi}} + 3 \widetilde{<x^2>_{\phi}}^2 \;\;\mbox{etc.}
\label{centralsimpler}
\end{equation}

The significance of the second term in the expression for $\mu_4$ can be understood in terms of the kurtosis $\kappa = \mu_4/\mu_2^2 - 3$. The kurtosis is a measure of the peakedness of a distribution. Higher $\kappa$ values result in a sharper peak with heavier tails. The normal distribution possesses $\kappa = 0$. From (\ref{centralsimpler}), one finds that, $\kappa = \widetilde{<x^4>_{\phi}}/\widetilde{<x^2>_{\phi}}^2$, which is always positive. Thus, the position probability for the memoryless random walk will always be more sharply peaked than the normal distribution. However, if $<x^4>_{\phi}$ and $<x^2>_{\phi}$ are time-independent, then, $\kappa$ will approach zero for large values of $t$.

Also, if the jump probability is symmetric such that $<x^n>_{\phi}$ vanishes for all odd values of $n$, then, from (\ref{moment_ml}) and (\ref{derivatives}), it is evident that all odd moments of $p(\vec{r},t)$ will vanish as well. Therefore, the position probability density will be symmetric, as expected.

\section{Results for General Random Walk}
Here we proceed to the more challenging task of deriving the moments for a general waiting time density. We start by expressing the moments of $p(\vec{r},t)$ in terms of the intermediate density $f(\vec{r},t)$ and waiting time density $\psi(t)$. Using (\ref{divide}) and (\ref{conv1}), one can write,
\begin{align}
\nonumber <x>(t) \;=&  \int_{-\infty}^{\infty} x p(\vec{r},t) d^3 \vec{r} \\
\nonumber =&  \int_{-\infty}^{\infty} x q(\vec{r},t) d^3 \vec{r} \\
\nonumber =& \;i \bar{q}'(\vec{0},t) \qquad\qquad\quad\mbox{using (\ref{moments})}\; .\\
=& \;i \bar{f}'(\vec{0},t) \ast \Psi(t) \qquad \mbox{where $\ast$ denotes convolution in time domain.} \label{meanx}
\end{align}

Taking the Laplace transform of both sides, one obtains, $\hat{<x>}(s) = i \hat{\bar{f}}'(\vec{0},s) \hat{\Psi}(s)$. But, according to the definition of $\Psi(t)$ in (\ref{Psi}), $\hat{\Psi}(s) = (1 - \hat{\psi}(s))/s$. Hence, we get,
\begin{equation}
\hat{<x>}(s) = i \hat{\bar{f}}'(\vec{0},s) \bigg(\frac{1 - \hat{\psi}(s)}{s}\bigg)\;.
\label{xlaplace}
\end{equation}

Similarly, one can show that,
\begin{eqnarray}
\hat{<x^2>}(s) &=& -\hat{\bar{f}}''(\vec{0},s) \bigg(\frac{1 - \hat{\psi}(s)}{s}\bigg)\;,
\label{xsqlaplace} \\
\mbox{and, in general,}\quad
\hat{<x^n>}(s) &=& i^n \hat{\bar{f}}^{(n)}(\vec{0},s) \bigg(\frac{1 - \hat{\psi}(s)}{s}\bigg)\;,\quad n=1,2,3,...\;.
\label{xnlaplace}
\end{eqnarray}

Therefore, in order to determine the moments of $p(\vec{r},t)$, we must first find out $\hat{\bar{f}}^{(n)}(\vec{0},s)$ from the integral equation (\ref{recursion_k}). For the purpose of simplifying mathematical notations, we define a linear operator $A$ by stating its action on an arbitrary function $g(\vec{k},t)$ as,
\begin{equation}
A [g(\vec{k},t)] \;\doteq \; \bar{\phi}(\vec{k}|t)\{g(\vec{k},t) \ast \psi(t)\}\; .
\label{A}
\end{equation}

Equation (\ref{recursion_k}) can then be written in a compact form,
\begin{equation}
\bar{f}(\vec{k},t) - A [\bar{f}(\vec{k},t)] = \bar{\phi}(\vec{k}|t) \psi(t)\; .
\label{elegant1}
\end{equation}

Here we note a property of the operator $A$. Consider any two functions $h(\vec{k},t)$ and $g(\vec{k},t)$ related by $h(\vec{k},t) = A [g(\vec{k},t)]$. Using (\ref{A}), one can write,
\begin{equation}
h'(\vec{k},t) = \frac{\bar{\phi}'(\vec{k}|t)}{\bar{\phi}(\vec{k}|t)} h(\vec{k},t) + A [g'(\vec{k},t)]\; .
\label{prop}
\end{equation}

Taking derivatives of both sides in (\ref{elegant1}) and applying the above property, we obtain,
\begin{eqnarray}
\nonumber &&\bar{f}'(\vec{k},t) - \frac{\bar{\phi}'(\vec{k}|t)}{\bar{\phi}(\vec{k}|t)} (\bar{f}(\vec{k},t) - \bar{\phi}(\vec{k}|t) \psi(t)) - A [\bar{f}'(\vec{k},t)] = \bar{\phi}'(\vec{k}|t) \psi(t) \\
\mbox{or,}\quad &&\bar{f}'(\vec{k},t) - A [\bar{f}'(\vec{k},t)] = \frac{\bar{\phi}'(\vec{k}|t)}{\bar{\phi}(\vec{k}|t)} \bar{f}(\vec{k},t)\; .
\label{elegant2}
\end{eqnarray}

Using (\ref{moments}) and the fact that $\phi(\vec{r}|t)$ is a probability density, one can write,
\begin{eqnarray}
\bar{\phi}(\vec{0}|t) &=& \int_{-\infty}^{\infty} \phi(\vec{r}|t) d^3 \vec{r} = 1 \; ,\label{phi0}\\
\mbox{and,}\quad\bar{\phi}'(\vec{0}|t) &=& \frac{1}{i} <x>_{\phi}(t)\; .\label{phid0}
\end{eqnarray}

Again, it follows from (\ref{A}) that,
\begin{equation}
A [g(\vec{k},t)]\big|_{\vec{k}=\vec{0}} = g(\vec{0},t) \ast \psi(t) \; .
\label{A0}
\end{equation}

Application of these results on (\ref{elegant1}) and (\ref{elegant2}) yields,
\begin{eqnarray}
\hat{\bar{f}}(\vec{0},s) &=& \frac{\hat{\psi}(s)}{1 - \hat{\psi}(s)} = \hat{M}_{\psi}(s) \quad\mbox{(say) .} \label{f0}\\
\mbox{and,}\quad \hat{\bar{f}}'(\vec{0},s) &=& \frac{\mathcal{L}\big[<x>_{\phi}(t) M_{\psi}(t)\big]}{i(1 - \hat{\psi}(s))}\; .\label{fd0}
\end{eqnarray}

Substitution of (\ref{fd0}) into (\ref{xlaplace}) gives,
\begin{equation}
\hat{<x>}(s) = \;\frac{\mathcal{L}\big[<x>_{\phi}(t) M_{\psi}(t)\big]}{s}\; .\label{xsols}
\end{equation}

Taking inverse Laplace transform, we get the general expression of the mean position,
\begin{equation}
<x>(t) = \int_{0}^{t} <x>_{\phi}(\tau) \;M_{\psi}(\tau) \;d \tau \; .
\label{solution1}
\end{equation}

Combining analogous expressions for $<y>$ and $<z>$ and making use of (\ref{f0}), we can thus write,
\begin{equation}
<\vec{r}>(t) = \int_{0}^{t} <\vec{r}>_{\phi}(\tau) \;\mathcal{L}^{-1} \bigg[\frac{\hat{\psi}(s)}{1 - \hat{\psi}(s)} \bigg] \;d \tau \; .
\label{solution1r}
\end{equation}

This can be explained as follows. From (\ref{njumps}), we find that the probability of exactly $n$ jumps taking place in duration $t$ can be written as,
\begin{equation}
\mathcal{P}_n(t) = \underbrace{\psi(t) \ast \psi(t) \ast \dots \ast \psi(t) \ast \Psi(t)}_\text{$(n + 1)$ convolutions}\; .
\label{njconv}
\end{equation}

Therefore, the mean number of jumps occurring in time $t$ is given by,
\begin{align}
\nonumber <N(t)> \;=& \;\mathcal{L}^{-1} \bigg[ \sum_{n=0}^{\infty} n \hat{\mathcal{P}}_n(s) \bigg] \\
\nonumber =& \;\mathcal{L}^{-1} \bigg[ \sum_{n=0}^{\infty} n (\hat{\psi}(s))^n \Big(\frac{1 - \hat{\psi}(s)}{s} \Big) \bigg] \\
=& \;\mathcal{L}^{-1} \bigg[ \frac{\hat{\psi}(s)}{s(1 - \hat{\psi}(s))} \bigg]\; .
\label{njlaplace}
\end{align}

Thus, on average, the number of steps taken by the random walker in an infinitesimal time interval $dt$ is,
\begin{equation}
\frac{d <N(t)>}{dt} dt = \mathcal{L}^{-1} \bigg[ \frac{\hat{\psi}(s)}{1 - \hat{\psi}(s)} \bigg] dt\; .
\label{njdt}
\end{equation}

Now we see that the integrand in (\ref{solution1r}) is the product of the average number of jumps in duration $d\tau$ and the average displacement of the walker in each such jump. So, it represents the average displacement in the interval $d\tau$, which when integrated, gives us the mean position of the random walker after time $t$, as expected. In fact, this reasoning was used to derive the linear response of systems to time-dependent external fields \cite{physica}.

For the special case of the memoryless random walk, $\psi(t) = (1/T) \exp(-t/T)$. Hence, $\hat{\psi}(s) = 1/(Ts + 1)$, giving $\hat{M_{\psi}}(s) = 1/Ts$ and $M_{\psi}(t) = 1/T$. Putting this into (\ref{solution1}), we get, $<x>(t) = \;(1/T) \int_{0}^{t} <x>_{\phi} (\tau) \;d\tau$, in agreement with what we had derived in (\ref{compact}).

Next we proceed to calculate the higher moments. Keeping analogy with (\ref{elegant1}) and (\ref{elegant2}), let us write,
\begin{equation}
\bar{f}^{(n)}(\vec{k},t) - A [\bar{f}^{(n)}(\vec{k},t)] = \bar{\chi}_n(\vec{k},t)\; ,\quad n=1,2,3,\dots \; .
\label{elegant3}
\end{equation}

Then (\ref{xnlaplace}) and (\ref{A0}) together imply,
\begin{eqnarray}
\hat{<x^n>}(s) &=& \frac{i^n}{s} \hat{\bar{\chi}}_n(\vec{0},s)\; , \label{xnchis} \\
\mbox{or,}\quad \frac{d}{dt} <x^n>(t) &=& i^n \bar{\chi}_n(\vec{0},t)\; .
\label{xnchi}
\end{eqnarray}

$\bar{\chi}_n(\vec{k},t)$ can be obtained by taking the $n$-th derivative of (\ref{elegant1}) and substituting the result in (\ref{elegant3}),
\begin{equation}
\bar{\chi}_n(\vec{k},t) = \frac{\partial^n}{\partial k_x^n} \big(\bar{\phi}(\vec{k}|t)\{\bar{f}(\vec{k},t) \ast \psi(t)\}\big) - \bar{\phi}(\vec{k}|t)\{\bar{f}^{(n)}(\vec{k},t) \ast \psi(t)\} + \bar{\phi}^{(n)}(\vec{k}|t) \psi(t)\; .
\label{leibnitz1}
\end{equation}

Expanding the first term by Leibnitz's formula, we get,
\begin{align}
\bar{\chi}_n(\vec{k},t) =& \;\bar{\phi}^{(n)}(\vec{k}|t) \psi(t) + \sum_{q=1}^{n} \binom{n}{q} \bar{\phi}^{(q)}(\vec{k}|t)\{\bar{f}^{(n-q)}(\vec{k},t) \ast \psi(t)\}
\label{leibnitz2} \\
=& \;\bar{\phi}^{(n)}(\vec{k}|t)\{\psi(t) + \bar{f}(\vec{k},t) \ast \psi(t) \} + \sum_{q=1}^{n-1} \binom{n}{q} \bar{\phi}^{(q)}(\vec{k}|t)\{\bar{f}^{(n-q)}(\vec{k},t) \ast \psi(t)\}\; .
\label{leibnitz3}
\end{align}

The term outside summation can be simplified for $\vec{k}=\vec{0}$ by noting that,
\begin{equation}
\mathcal{L} [\psi(t) + \bar{f}(\vec{0},t) \ast \psi(t)] = \hat{\psi}(s) (1 + \hat{M}_{\psi}(s)) = \hat{M}_{\psi}(s) \quad \mbox{[using (\ref{f0})]}\; .
\label{simplify}
\end{equation}

This result, along with (\ref{moments}), (\ref{xnlaplace}) and (\ref{xnchi}), yield the following equation giving the $n$-th moment in terms of the lower moments, $(n = 2,3,4,\dots)$,
\begin{align}
\frac{d}{dt} <x^n>(t) =& <x^n>_{\phi}(t) M_{\psi}(t) + \sum_{q=1}^{n-1} \binom{n}{q} <x^q>_{\phi}(t) \mathcal{L}^{-1} \big[s \;\hat{<x^{n-q}>}(s) \hat{M}_{\psi}(s) \big] \label{main1}\\
=& <x^n>_{\phi}(t) M_{\psi}(t) + \sum_{q=1}^{n-1} \binom{n}{q} <x^q>_{\phi}(t) \Big\{ M_{\psi}(t) \ast \frac{d}{dt} <x^{n-q}>(t) \Big\}\; .\label{main}
\end{align}

For the memoryless random walk, we noted that $<x^n>$ depended on $<x^m>_{\phi}$ if and only if $m \leq n$. According to the equation above, this fact remains true in the general case. It can be readily verified that (\ref{main}) reproduces the results for the memoryless random walk for the choice $\psi(t) = (1/T) \exp(-t/T)$. As we found earlier, for this waiting time density, $M_{\psi}(t) = 1/T$. Substituting this in (\ref{main}), we obtain,
\begin{equation}
\frac{d}{dt} <x^n>(t) = \frac{1}{T} \sum_{q=1}^{n} \binom{n}{q} <x^q>_{\phi}(t) <x^{n-q}>(t) \; ,
\label{reduction1}
\end{equation}
with the understanding that $<x^0>(t) = 1$. That this equation is indeed equivalent to (\ref{moment_ml}) can be confirmed as follows. Differentiating (\ref{moment_ml}) with respect to $t$ and using (\ref{g}), one can write,
\begin{align}
\nonumber \frac{d}{dt} <x^n>(t) =& -\frac{1}{T} <x^n>(t) + \frac{1}{T} i^n e^{-t/T} \frac{\partial^n}{\partial k_x^n} \Big[ e^{g(\vec{k},t)} \bar{\phi} (\vec{k},t) \Big] \bigg|_{\vec{k}=\vec{0}} \\
\nonumber =& \;\frac{1}{T} \bigg[ -<x^n>(t) + \sum_{q=0}^{n} \binom{n}{q} \big( i^q \bar{\phi}^{(q)}(\vec{0},t) \big) \bigg(i^{n-q} e^{-t/T} \frac{\partial^{n-q}}{\partial k_x^{n-q}} e^{g(\vec{k},t)} \Big|_{\vec{k} = \vec{0}} \bigg) \bigg] \\
=& \;\frac{1}{T} \sum_{q=1}^{n} \binom{n}{q} <x^q>_{\phi}(t) <x^{n-q}>(t) \; , \label{reduction2}
\end{align}
which is identical to (\ref{reduction1}).

Equation (\ref{main}) and (\ref{solution1}) constitute the complete solution for the moments $<x^n>\;(n \in \mathcal{Z}^+)$ of $p(\vec{r},t)$. These equations, along with their counterparts for $y$ and $z$, completely specify the random walker's position probability density $p(\vec{r},t)$.

As a special case, let us investigate the standard deviation $\Delta x$. For $n=2$, (\ref{main}) becomes, after substituting from (\ref{solution1}),
\begin{equation}
\frac{d}{dt} <x^2>(t) = \;<x^2>_{\phi}(t) M_{\psi}(t) + 2 <x>_{\phi}(t)\; \big\{M_{\psi}(t) \ast (<x>_{\phi}(t) M_{\psi}(t))\big\}\; .
\label{presolution2}
\end{equation}

Hence,
\begin{align}
\nonumber \frac{d}{dt} <(\Delta x)^2>(t) =& \frac{d}{dt} <x^2>(t) - 2 <x>(t) \;\frac{d}{dt} <x>(t) \\
\nonumber =& \;<x^2>_{\phi}(t) M_{\psi}(t) + 2 <x>_{\phi}(t)\;\Big[ \big\{M_{\psi}(t) \ast (<x>_{\phi}(t) M_{\psi}(t))\big\} \\
&- M_{\psi}(t) \int_{0}^{t} <x>_{\phi}(\tau) M_{\psi}(\tau) \;d \tau \Big]\; .\label{solution2}
\end{align}

Thus, the standard deviation is given by,
\begin{align}
\nonumber \Delta x =& \sqrt{\int_{0}^{t} <x^2>_{\phi}(\tau) M_{\psi}(\tau)\;d\tau +  \int_{0}^{t} 2 <x>_{\phi}(\tau_1)\;\Big[ \big\{M_{\psi}(\tau_1) \ast (<x>_{\phi}(\tau_1) M_{\psi}(\tau_1))\big\}} \\
& \hspace{6cm} - M_{\psi}(\tau_1) \int_{0}^{\tau_1} <x>_{\phi}(\tau_2) M_{\psi}(\tau_2) \;d \tau_2 \Big] \;d\tau_1\; .
\label{solution3}
\end{align}

If the random walk is memoryless, $M_{\psi}(t) = 1/T$, a constant. As a result, the second part in the above equation vanishes and we get,
\begin{equation}
(\Delta x)_{\mbox{memoryless}} = \sqrt{\frac{1}{T} \int_{0}^{t} <x^2>_{\phi}(\tau)\; d\tau} \; ,
\label{mlreduce}
\end{equation}
conforming to what we found earlier (see (\ref{compact})).

\section{Discussion of Results}

\subsection{A Case Study : Failure of generalized Fokker-Planck equation in predicting higher moments}
A few years ago, Sokolov and Klafter\cite{PRL} analyzed a similar problem where the random walk is one-dimensional and the walker takes steps of unit length in either direction. This is seen to be a special case of our problem where the jump probability is given by,
\begin{equation}
\phi(\vec{r}|t) = \bigg(\frac{1}{2} + \frac{\mu}{2} f(t)\bigg) \delta(\vec{r} - \vec{a}) + \bigg(\frac{1}{2} - \frac{\mu}{2} f(t)\bigg) \delta(\vec{r} + \vec{a})\; .
\label{Prl}
\end{equation}
where $\vec{a}$ is a unit vector along positive $x$ direction. Consequently, the moments are given by,
\begin{align}
<x^n>_{\phi} \;=&\; \mu f(t) \qquad\;\mbox{if $n$ is odd ,} \label{oddmoments}\\
=&\; 1 \qquad\qquad\mbox{if $n$ is even .}
\label{evenmoments}
\end{align}

When these are substituted, Eqs. (\ref{solution1}) and (\ref{presolution2}) indeed reproduce Eqs. (12) and (13) in Ref. \cite{PRL}. Thus, the expressions we deduced for the mean position and standard deviation of the random walker agree with those in Ref. \cite{PRL} found via a different method. Consequently, the comments made there regarding the phenomena of ``field-induced dispersion'' and ``death of linear response" in systems showing sub-diffusion apply here also. However, the higher moments obey the following equation, obtained after substituting (\ref{oddmoments}) and (\ref{evenmoments}) into (\ref{main1}),
\begin{equation}
\frac{d}{dt} <x^n>(t) = \sum_{q=2,4,6,\dots}^{} \binom{n}{q} \hat{\Phi} <x^{n-q}>(t) + \mu f(t) \sum_{q=1,3,5,\dots}^{} \binom{n}{q} \hat{\Phi} <x^{n-q}>(t) \; .
\label{highermoments}
\end{equation}
where $\hat{\Phi} <x^{n-q}>(t) \equiv \mathcal{L}^{-1} \big[s \;\hat{<x^{n-q}>}(s) \hat{M}_{\psi}(s) \big]$ coincides with the integro-differential operator used in Ref. \cite{PRL}. The general equation for moments derived in Ref. \cite{PRL} (see equation (11)) from a generalized Fokker-Planck equation (FPE) contains only the first two terms corresponding to $q=1$ and $q=2$ in (\ref{highermoments}). Similar equations were derived in Refs. \cite{heinsalu1, heinsalu2, PRE}. The two equations match only if we substitute $<x^n>_{\phi}\; =\; 0\; \forall \; n \geq 3$ (which is not possible for any choice of $\phi(\vec{r}|t)$).
However, we note that the loss of the additional terms is an artifact of the approximation involved in taking the continuum limit of the Generalized Master Equation (GME) to derive the generalized FPE. As illustrated in Ref. \cite{PRL}, application of probability conservation arguments yields the following GME,
\begin{equation}
\dot{p}_k(t) = \bigg(\frac{1}{2} + \frac{\mu}{2} f(t) \bigg) \hat{\Phi} p_{k-1}(t) + \bigg(\frac{1}{2} - \frac{\mu}{2} f(t) \bigg) \hat{\Phi} p_{k+1}(t) - \hat{\Phi} p_{k}(t)\; ,
\label{GME}
\end{equation}
where $p_k(t)$ denotes the probability that the random walker is at site $k$ at time $t$. The $n$-th moment can be calculated as, $<x^n>(t) = \sum_{k=-\infty}^{\infty} k^n p_k(t)$. Thus, multiplying (\ref{GME}) by $k^n$ and summing over $k$, we get,
\begin{align}
\nonumber \frac{d}{dt} <x^n>(t) =& \; - \hat{\Phi} <x^n>(t) + \bigg(\frac{1}{2} + \frac{\mu}{2} f(t) \bigg) \hat{\Phi} \big[\sum_{k=-\infty}^{\infty} (k+1)^n p_k(t)\big] \\
\nonumber & \;+ \bigg(\frac{1}{2} - \frac{\mu}{2} f(t) \bigg) \hat{\Phi} \big[\sum_{k=-\infty}^{\infty} (k-1)^n p_k(t)\big] \\
\nonumber =& \;- \hat{\Phi} <x^n>(t) +  \bigg(\frac{1}{2} + \frac{\mu}{2} f(t) \bigg) \hat{\Phi} \big[\sum_{q=0}^{n} \binom{n}{q} <x^{n-q}>(t)\big] \\
\nonumber & \;+ \bigg(\frac{1}{2} - \frac{\mu}{2} f(t) \bigg) \hat{\Phi} \big[\sum_{q=0}^{n} (-1)^q \binom{n}{q} <x^{n-q}>(t)\big]  \\
=& \sum_{q=2,4,6,\dots}^{} \binom{n}{q} \hat{\Phi} <x^{n-q}>(t) + \mu f(t) \sum_{q=1,3,5,\dots}^{} \binom{n}{q} \hat{\Phi} <x^{n-q}>(t) \; ,
\label{GMEderivation}
\end{align}
which is identical to (\ref{highermoments}). This clearly shows that the approximations involved in taking the continuum limit lead to the disappearance of these additional terms. Although this does not affect the mean and standard deviation, the higher moments are increasingly affected. 

The significance of the additional terms in our analysis may be investigated by examining how they modify the asymmetry and peakedness of the position probability density $p(\vec{r},t)$. For example, the 3rd central moment, given by $\mu_3 = \;<(x - <x>)^3>$, is a measure of the lopsidedness of a distribution. This can be expanded to give, $\mu_3 = \;<x^3> - 3 <x^2><x> + 2 <x>^3$. Thus, the extra term in (\ref{highermoments}) for $n=3$ increases the 3rd central moment by an amount $\Delta \mu_3 = \mu \int_{0}^{t} f(\tau) M_{\psi}(\tau) d\tau$. A quick reflection reveals that this is just the expression for the mean position $<x>$. The skewness $\gamma$ is defined by, $\gamma = \mu_3 / (\Delta x)^3$. Consequently, the additional term results in an increment of $\gamma$ given by,
\begin{equation}
\Delta \gamma = \frac{<x>}{(\Delta x)^3}\; .
\label{assymchange}
\end{equation}

A positive skewness implies that the probability density function has a longer tail to the right of the mean so that the bulk of the values lie to its left and vice-versa. Similarly, a measure of the peakedness of a probability distribution is the kurtosis defined as, $\kappa = \;\mu_4/ (\Delta x)^4 - 3$, where $\mu_4$ is the 4th central moment, given by, $\mu_4 = <(x - <x>)^4>$. Higher value of $\kappa$ indicates that the distribution has a more acute peak around the mean and fatter tails. Therefore, more of the variance is the result of infrequent extreme deviations, as opposed to frequent but modest deviations. The additional terms in (\ref{highermoments}) increases the kurtosis by,
\begin{equation}
\Delta \kappa = \frac{2 <x^2> - 4 <x>^2 - \int_{0}^{t} M_{\psi}(\tau) d\tau}{(\Delta x)^4}\; .
\label{peakchange}
\end{equation}

Let us examine the order of magnitudes of these changes for the particular case of oscillating external field $f(t) = f_0 \sin \omega t$ and subdiffusive waiting-time density ($\psi(t) \sim t^{-1-\alpha}$ with $0 < \alpha < 1$) investigated in Ref. \cite{PRL}. By using the shift and convolution properties of Laplace transforms and taking limit as $s\to 0$, one can show that for large values of $t$, $<x> \;\simeq \mu f_0 \;\mbox{Im} \{\hat{M}_{\psi}(-i\omega)\}$, a constant and $\Delta \mu_4$ increases as $t^\alpha$ (see (\ref{meanfinal}) and (\ref{changemu4})). Again, From (\ref{meanfinal}) and (\ref{squarefinal}), we see that for large values of $t$, the variance behaves as,
\begin{equation}
(\Delta x)^2 \simeq \big(1 + \mu^2 f_0^2 \mbox{Re}\{z_1\}\big) \int_{0}^{t} M_{\psi}(\tau) d\tau \;- \mu^2 f_0^2 \mbox{Re}\{z_1 z_2\} - \mu^2 f_0^2 \;\big(\mbox{Im} \{z_1\}\big)^2\; ,
\label{variance}
\end{equation}
which increases as $t^\alpha$. Substitution of these results in (\ref{assymchange}) and (\ref{peakchange}) yields $\Delta \gamma \sim t^{-3\alpha/2}$ and $\Delta \kappa \sim t^{-\alpha}$. However, since we are concerned in how these additional terms modify the results in Ref. \cite{PRL}, the quantities of interest are not the absolute increments, rather the relative increments, $\Delta \gamma / \gamma$ and $\Delta \kappa / \kappa$. As we show in the Appendices (see (\ref{relgammacalculation}) and (\ref{relkappacalculation})), for large values of $t$,
\begin{align}
\frac{\Delta \gamma}{\gamma} &\simeq \frac{\mbox{Im}\{z_1\}}{6 \mbox{Re}\{z_1\} \mbox{Im}\{z_1\} -\frac{3}{2} \mu^2 f_0^2 \mbox{Im}\{z_1 z_2 (z_3 - z_1) \}  + 3 \mu^2 f_0^2 \mbox{Im}\{z_1 \} \mbox{Re}\{z_1 z_2\} + 2 \mu^2 f_0^2 \big(\mbox{Im}\{z_1 \}\big)^3} \;\; \label{relativegamma} \\
&\mbox{and,}\quad\frac{\Delta \kappa}{\kappa} \simeq \frac{\big(1 + 4\mu^2 f_0^2 \mbox{Re}\{z_1\} \big) \int_{0}^{t} M_{\psi}(\tau) d\tau + C}{6\big(1 + \mu^2 f_0^2 \mbox{Re}\{z_1\}\big)^2 \int_{0}^{t} M_{\psi}(\tau) \ast M_{\psi}(\tau) d\tau + A \int_{0}^{t} M_{\psi}(\tau) d\tau + B}\; .
\label{relativekappa}
\end{align}
where $z_1 = \hat{M}_{\psi}(-i \omega)$, $z_2 = \hat{M}_{\psi}(-i 2\omega)$, $z_3 = \hat{M}_{\psi}(-i 3\omega)$, $z_4 = \hat{M}_{\psi}(-i 4\omega)$ and $A$, $B$, $C$ are constants related to $z_i \; (i=1,2,3,4)$. For subdiffusive waiting time density $\psi(t) \sim t^{-1-\alpha}$ with $0 < \alpha < 1$, $\int_{0}^{t} M_{\psi}(\tau) d\tau \sim t^\alpha$ and $\int_{0}^{t} M_{\psi}(\tau) \ast M_{\psi}(\tau) d\tau \sim t^{2\alpha}$. Therefore, for large values of $t$, $\Delta \gamma / \gamma$ approaches a constant and $\Delta \kappa / \kappa \sim t^{-\alpha}$. Thus, in terms of their effect on peakedness of the position probability density, the additional terms become more significant for smaller $\alpha$.

It is also of interest to find out what value $\gamma$ and $\kappa$ assume for large values of $t$. This can be useful to determine to what extent the position probability of the random walker differs from the normal distribution, which has both $\gamma = 0$ and $\kappa = 0$. First, we have already noted that $\Delta \gamma \sim t^{-3\alpha/2}$ and  $\Delta \gamma / \gamma$ approaches a constant. Therefore, we conclude that $\gamma$ must decay towards zero as $t^{-3\alpha/2}$. However, both $\Delta \kappa / \kappa$ and $\Delta \kappa$ decrease as $t^{-\alpha}$. Thus, for large $t$, the kurtosis approaches a constant. From (\ref{bigcal4}) and (\ref{variance}), one finds that this constant is given by,
\begin{equation}
\kappa \simeq 6\;\frac{\int_{0}^{t} M_{\psi}(\tau) \ast M_{\psi}(\tau) d\tau}{\big(\int_{0}^{t} M_{\psi}(\tau) d\tau\big)^2} - 3\; .
\label{kappaconst}
\end{equation}

Let us take $M_{\psi}(t) = t^{\alpha - 1}$, corresponding to subdiffusive waiting time densities. Then, $\hat{M}_{\psi}(s) = \Gamma(\alpha)/s^\alpha$. Therefore, $M_{\psi}(t) \ast M_{\psi}(t) = \mathcal{L}^{-1} [(\hat{M}_{\psi}(s))^2] = ((\Gamma(\alpha))^2/\Gamma(2\alpha)) t^{2\alpha - 1}$. Substituting these into (\ref{kappaconst}), we obtain,
\begin{equation}
\kappa \simeq 3\bigg[\alpha \frac{(\Gamma(\alpha))^2}{\Gamma(2\alpha)} - 1\bigg]\; .
\label{kappaalpha}
\end{equation}

\begin{figure}[h]
\centering
{\includegraphics[width=8cm]{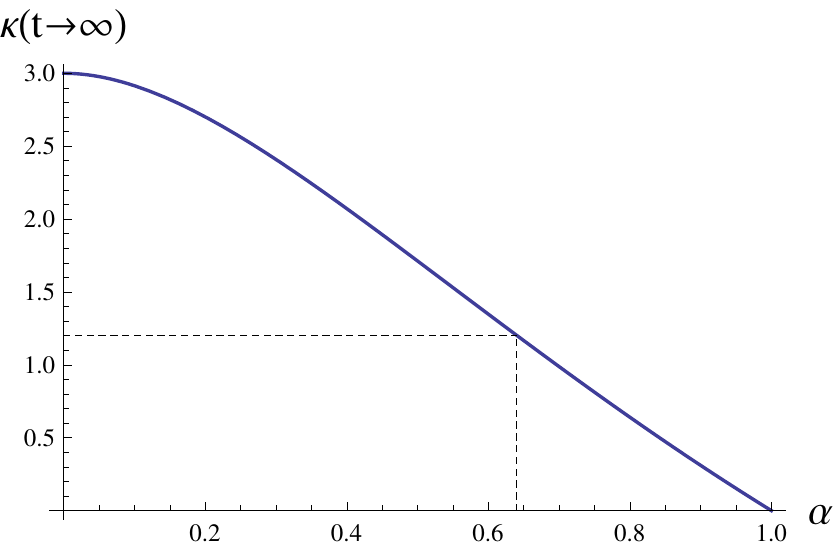}}
\caption{Variation of kurtosis for large $t$ with $\alpha$.}
\label{kurtosisfig}
\end{figure}

Figure \ref{kurtosisfig} plots this limiting value of $\kappa$ as a function of the exponent $\alpha$. For $\alpha = 1$, $\kappa(t \to \infty) = 0$. This is expected because we know that $\alpha = 1$ reproduces normal diffusion for which the position probability of the particle approaches a Gaussian distribution for large $t$. The figure depicts that as we lower $\alpha$ towards zero, $\kappa(t \to \infty)$ rises monotonically from $0$ towards $3$. This means that $p(\vec{r},t \to \infty)$ becomes more and more sharply peaked with fatter tails. The probability of infrequent large deviations from the mean increases while that of frequent modestly sized deviations decreases. This clearly shows how the normal approximation breaks down more and more as one enters deeper into the subdiffusive regime. The figure also provides us a good picture of the position probability density in the large time limit in terms of well-known distributions for certain values of $\alpha$. For example, the logistic distribution has zero skewness and a kurtosis of $1.2$. From Figure \ref{kurtosisfig}, we see that $\kappa(t \to \infty) = 1.2$ for $\alpha \approx 0.64$. Thus, we can say that for $\alpha \approx 0.64$, the pdf of the logistic distribution with mean and variance given by (\ref{meanfinal}) and (\ref{variance}) respectively provides a good approximation to $p(\vec{r},t \to \infty)$. It describes accurately the first four central moments and since the higher moments generally measure properties further and further out in the tails, we can get a fair idea of its overall shape, especially near the mean. Similarly, $p(\vec{r},t \to \infty)$ is well approximated by the pdf of the Laplace distribution for $\alpha \to 0$.

Now consider the probability density function
\begin{equation}
f_{\beta}(x|\mu,b) = \frac{\beta}{2 b^{1/\beta} \Gamma(1/\beta)} \exp\bigg(-\frac{|x - \mu|^\beta}{b}\bigg)\; .
\label{beta}
\end{equation}

It has the following characterizations,
\begin{align}
<x> \;=&\; \mu \; ,\qquad
(\Delta x)^2 \;=\; b^{2/\beta} \frac{\Gamma(3/\beta)}{\Gamma(1/\beta)} \; ,\label{betameanvariance}\\
\gamma \;=&\; 0 \; ,\qquad
\kappa \;=\; \frac{\Gamma(5/\beta) \Gamma(1/\beta)}{(\Gamma(3/\beta))^2} - 3\; .\label{betagammakappa}
\end{align}

\begin{figure}[h]
\centering
{\includegraphics[width=8cm]{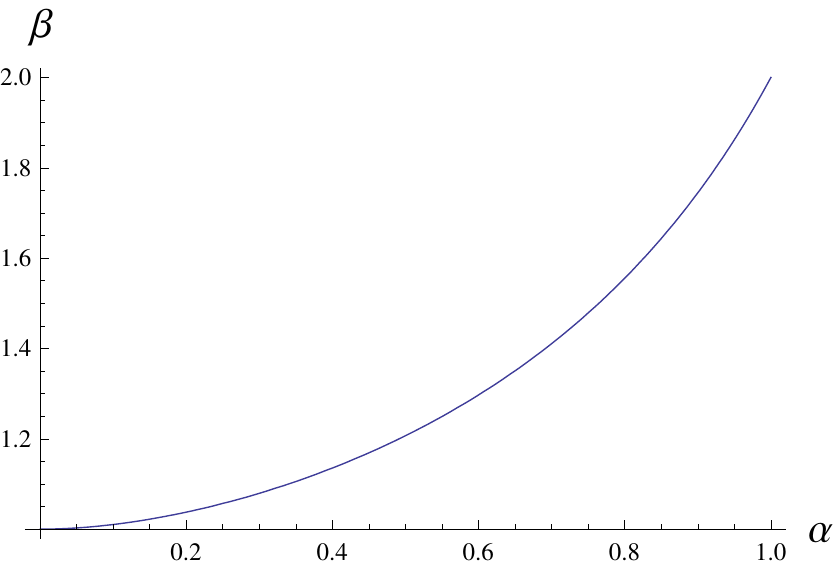}}
\caption{$\beta$ vs $\alpha$: The curve is well approximated by $\beta = \alpha^{2.4} + 1$.}
\label{betaalphafig}
\end{figure}

$\kappa$ continuously increases from $0$ to $3$ as the parameter $\beta$ decreases from $2$ to $1$. Therefore, for any value of $\alpha \in (0,1)$, there exists a unique value of $\beta \in (1,2)$, such that $f_{\beta}(x|\mu,b)$ has the same kurtosis as $p(\vec{r},t \to \infty)$. This is illustrated in Figure \ref{betaalphafig}.

Thus, for suitable choices of $\mu$ and $b$ in (\ref{beta}), we can get a rough picture of $p(\vec{r},t \to \infty)$ for any $\alpha \in (0,1)$.

\subsection{Other Issues: Choice of the jump probability density}

The particular form of the jump probability density given by (\ref{Prl}), assumed in recent works, implies that $<x^n>_{\phi}$ is independent of the external field for all even values of $n$. This may not be true in all circumstances, e.g., when the external field is temperature. An increase in temperature is not expected to make a jump towards right more probable than a jump towards left. Rather, the increased thermal energy of the particle should allow it to hop longer distances, thus increasing $<x^2>_{\phi}$. This can arise, for example, if a rise in temperature increases the width of $\phi(\vec{r}|t)$ which has the shape of a Gaussian distribution centered at the origin. In such a circumstance, $<x>_{\phi} = 0$. Therefore, from (\ref{solution1}), $<x> = 0$. However, $<x^2>_{\phi}(t) = g(T(t))$, where $g$ is an increasing function of temperature $T$. Hence, from (\ref{presolution2}), $<x^2>(t) = \int_{0}^{t} g(T(\tau)) M_{\psi}(\tau) d\tau$. If, for instance, $g$ is a linear function of $T$ which in turn rises linearly with time, then the standard deviation of the particle will increase as $t^{(\alpha + 1)/2}$ for subdiffusive waiting time densities, although its average displacement will be zero.

Further, the field-induced component of dispersion in the case of periodic zero-mean external field will, in general, depend on $\phi(\vec{r}|t)$. For example, if $\phi(\vec{r}|t) = (1/\sqrt{2 \pi}) \exp(-(x-\mu f(t))^2/2)$, then, $<x>_{\phi} = \mu f(t)$ and $<x^2>_{\phi}(t) = 1 + \mu^2 (f(t))^2$. Using (\ref{presolution2}), one finds that the dependence of $<x^2>_{\phi}(t)$ on $f(t)$ enhances the field-induced dispersion by an amount $\int_{0}^{t} \mu^2 (f(\tau))^2 M_{\psi}(\tau) d\tau$. For sinusoidally oscillating field $f(t)$, this contribution grows as $t^{\alpha}$ for large values of $t$, which is at the same scale as the original estimation in Ref. \cite{PRL} based on a nearest-neighbour jump model. Thus, relaxing the assumption of nearest neighbour jumps can significantly alter the predictions of the model.

\section{Concluding Remarks}
In this paper, we derived rigorous results for a general CTRW in three dimensions under time-dependent external forces. The method is based solely on probability arguments and does not invoke either the FFPE or the SLE. In this sense, the approach is more direct and illuminating. In the special case of CTRW on a one-dimensional lattice with nearest neighbour jumps, our equations confirm the phenomena of ``death of linear response" and ``field-induced dispersion" for zero-mean periodic forcing with subdiffusive waiting time densities. Our analysis illustrates the inability of the FFPE to correctly predict the higher order moments. We demonstrate how this discrepancy creeps in while taking the continuum limit from the GME. For sinusoidally varying fields, the effects of the additional terms on the skewness and kurtosis of the position probability density were studied. We also discussed how the predictions of the model can alter for different choices of the jump probability density and possible situations where these become relevant.

However, whether the results correspond to physical reality or not would depend on the extent to which the decoupled CTRW model itself can describe natural phenomena. Possible areas of concern include non-ergodicity \cite{nonergodic}, dependence of the waiting time statistics on the external field \cite{heinsalu2}, non-negligible travel time in jumps etc. Although most recent works concentrate on the integral moments of the position probability density, it has been suggested that fractional moments can give important information about its scaling behavior \cite{statmech}. It would also be interesting to find whether the direct probabilistic approach can be extended to incorporate both space and time dependence of external fields. Thus the CTRW formalism, which has been successful in modelling many natural phenomena, still requires plenty of theoretical and experimental research. 

\appendix
\section{Calculation of $\Delta \gamma / \gamma$}
\label{calculation1}
For an oscillating external field given by, $f(t) = f_0 \sin \omega t$, (\ref{solution1}) and (\ref{oddmoments}) give,
\begin{equation}
<x> (t) = \frac{\mu f_0}{2i} \int_{0}^{t} (e^{i\omega \tau} - e^{-i\omega \tau}) M_{\psi}(\tau) d\tau \; .
\label{mean1}
\end{equation}

Taking Laplace transform and applying shifting property, one obtains,
\begin{equation}
\hat{<x>}(s) = \frac{\mu f_0}{2is} \{\hat{M}_{\psi}(s - i\omega) - \hat{M}_{\psi}(s + i\omega)\}\; .
\label{mean2}
\end{equation}

For $s\to 0$ (corresponding to $t \to \infty$), this becomes,
\begin{equation}
\hat{<x>}(s) \simeq \frac{1}{s} \mu f_0\; \mbox{Im} \{\hat{M}_{\psi}(-i\omega)\} \quad\mbox{or,}\quad <x>(t) \simeq \mu f_0 \;\mbox{Im} \{\hat{M}_{\psi}(-i\omega)\}\; .
\label{meanfinal}
\end{equation}

For $n=2$, (\ref{highermoments}) yields,
\begin{equation}
<x^2>(t) = \int_{0}^{t} M_{\psi}(\tau) d\tau + 2 \mu f_0 \int_{0}^{t} \sin \omega \tau \frac{d}{d\tau} \int_{0}^{\tau} M_{\psi}(\tau - \tau') <x>(\tau') d\tau' d\tau \; .
\label{square1}
\end{equation}

Application of the shifting and convolution property of Laplace transforms on the above equation gives,
\begin{equation}
\hat{<x^2>}(s) = \frac{\hat{M}_{\psi}(s)}{s} + \frac{\mu f_0}{is}\big(\hat{M}_{\psi}(s - i\omega) \hat{<x>}(s - i\omega) - \hat{M}_{\psi}(s + i\omega) \hat{<x>}(s + i\omega)\big) \; .
\label{square2}
\end{equation}

Using (\ref{meanfinal}), we get, for $s \to 0$,
\begin{equation}
\hat{<x^2>}(s) \simeq \frac{\hat{M}_{\psi}(s)}{s} + \mu^2 f_0^2 \mbox{Re}\{\hat{M}_{\psi}(-i\omega)\} \frac{\hat{M}_{\psi}(s)}{s} - \frac{1}{s} \mu^2 f_0^2 \mbox{Re}\{\hat{M}_{\psi}(-i\omega) \hat{M}_{\psi}(-i2\omega)\} \; .
\label{square3}
\end{equation}

Hence, for large $t$,
\begin{equation}
<x^2>(t) \simeq \big(1 + \mu^2 f_0^2 \mbox{Re}\{\hat{M}_{\psi}(-i\omega)\} \big) \int_{0}^{t} M_{\psi}(\tau) d\tau \;- \mu^2 f_0^2 \mbox{Re}\{\hat{M}_{\psi}(-i\omega) \hat{M}_{\psi}(-i2\omega)\} \; .
\label{squarefinal}
\end{equation}

For subdiffusive waiting time densities $M_{\psi}(t) \sim t^{\alpha - 1}$, which means that the first term in (\ref{squarefinal}) increases as $t^\alpha$ and dominates the behavior of $<x^2>(t)$. 

Substituting $n=3$ in (\ref{highermoments}), one arrives at the following equation for $<x^3>$,
\begin{align}
\nonumber <x^3>(t) = \mu f_0 \int_{0}^{t} \sin\omega\tau M_{\psi}(\tau) d\tau \;&+\; 3 \mu f_0 \int_{0}^{t} \sin \omega \tau \frac{d}{d\tau} \int_{0}^{\tau} M_{\psi}(\tau - \tau') <x^2>(\tau') d\tau' d\tau \\
& + \;3 \int_{0}^{t} M_{\psi}(t - \tau) <x>(\tau) d\tau \; .
\label{cube1}
\end{align}

We want to analyze the effect of the first term in (\ref{cube1}) on the skewness. Comparison with (\ref{mean1}) shows that this is identical to $<x>$. Thus, according to (\ref{meanfinal}), for large value of $t$, this approaches the constant $\mu f_0 \;\mbox{Im} \{\hat{M}_{\psi}(-i\omega)\}$.

The 3rd term in (\ref{cube1}) can be written as,
\begin{align}
\nonumber 3 \int_{0}^{t} M_{\psi}(t - \tau) <x>(\tau) d\tau =&\; 3\; \mathcal{L}^{-1}\big[\hat{M}_{\psi}(s) \hat{<x>}(s)\big] \\
\nonumber =&\; \frac{3\mu f_0}{2i}\; \mathcal{L}^{-1}\bigg[\big(\hat{M}_{\psi}(s - i\omega) - \hat{M}_{\psi}(s + i\omega)\big) \frac{\hat{M}_{\psi}(s)}{s}\bigg]\; ,
\end{align}
which, for $s \to 0$ becomes $3\mu f_0 \mbox{Im} \{\hat{M}_{\psi}(-i\omega)\} \int_{0}^{t} M_{\psi}(\tau) d\tau$, and increases like $t^\alpha$ for large values of $t$. 

Let us denote the 2nd term in (\ref{cube1}) by $a(t)$. In order to find out how this behaves for large $t$, we can proceed in the same way as we did in going from (\ref{square1}) to (\ref{squarefinal}). This gives,
\begin{align}
\nonumber a(t) \simeq& \;3 \mu^3 f_0^3 \mbox{Im} \{\hat{M}_{\psi}(-i\omega)\} \mbox{Re} \{\hat{M}_{\psi}(-i\omega)\} \int_{0}^{t} M_{\psi}(\tau) d\tau \;+\; 6 \mu f_0 \mbox{Im} \{\hat{M}_{\psi}(-i\omega)\} \mbox{Re} \{\hat{M}_{\psi}(-i\omega)\} \\
&\;-\; \frac{3}{2} \mu^3 f_0^3 \mbox{Im} \big\{\hat{M}_{\psi}(-i\omega) \hat{M}_{\psi}(-i2\omega) \big(\hat{M}_{\psi}(-i3\omega) - \hat{M}_{\psi}(-i\omega)\big) \big\} \; .
\label{cube2}
\end{align}

Taking the contribution of only the second and third term, one can write,
\begin{align}
\nonumber <x^3>(t) \simeq& \;3\mu f_0 \mbox{Im} \{z_1\} (1 + \mu^2 f_0^2 \mbox{Re} \{z_1\}) \int_{0}^{t} M_{\psi}(\tau) d\tau \;+\; 6 \mu f_0 \mbox{Im} \{z_1\} \mbox{Re} \{z_1\} \\
&\;-\; \frac{3}{2} \mu^3 f_0^3 \mbox{Im} \{z_1 z_2 (z_3 - z_1)\} \; .
\label{cubefinal}
\end{align}
where $z_1 = \hat{M}_{\psi}(-i \omega)$, $z_2 = \hat{M}_{\psi}(-i 2\omega)$ and $z_3 = \hat{M}_{\psi}(-i 3\omega)$. Combining (\ref{meanfinal}), (\ref{squarefinal}) and (\ref{cubefinal}), one obtains, for the 3rd central moment,
\begin{align}
\nonumber \mu_3 =& <x^3> - 3 <x^2><x> + 2 <x>^3 \\
\simeq& \;6 \mu f_0 \mbox{Re}\{z_1\} \mbox{Im}\{z_1\} -\frac{3}{2} \mu^3 f_0^3 \mbox{Im}\{z_1 z_2 (z_3 - z_1) \}  + 3 \mu^3 f_0^3 \mbox{Im}\{z_1 \} \mbox{Re}\{z_1 z_2\} + 2 \mu^3 f_0^3 \big(\mbox{Im}\{z_1 \}\big)^3\; .
\label{mu3calculation}
\end{align}

The additional term in (\ref{cube1}), namely the first term, increases $\mu_3$ by an amount, $\Delta \mu_3 = \mu f_0 \mbox{Im} \{z_1\}$. Since the standard deviation remains unaffected, the relative increment of skewness is given by,
\begin{align}
\nonumber \frac{\Delta \gamma}{\gamma} =& \; \frac{\Delta \mu_3}{\mu_3} \\
\simeq&\; \frac{\mbox{Im}\{z_1\}}{6 \mbox{Re}\{z_1\} \mbox{Im}\{z_1\} -\frac{3}{2} \mu^2 f_0^2 \mbox{Im}\{z_1 z_2 (z_3 - z_1)\}  + 3 \mu^2 f_0^2 \mbox{Im}\{z_1 \} \mbox{Re}\{z_1 z_2\} + 2 \mu^2 f_0^2 \big(\mbox{Im}\{z_1 \}\big)^3}\; .
\label{relgammacalculation}
\end{align}

Thus, the relative increment of skewness approaches a constant for large $t$.

\section{Calculation of $\Delta \kappa / \kappa$}
\label{calculation2}
For calculating the 4th central moment, we need to evaluate $<x^4>$, which is governed, according to (\ref{highermoments}), by the following equation,
\begin{align}
\nonumber <x^4>(t) =& \int_{0}^{t} M_{\psi}(\tau) d\tau + 4 \mu f_0 \int_{0}^{t} \sin \omega \tau \frac{d}{d\tau} \int_{0}^{\tau} M_{\psi}(\tau - \tau') <x>(\tau') d\tau' d\tau \\
& + 6 \int_{0}^{t} M_{\psi}(t - \tau) <x^2>(\tau) d\tau + 4 \mu f_0 \int_{0}^{t} \sin \omega \tau \frac{d}{d\tau} \int_{0}^{\tau} M_{\psi}(\tau - \tau') <x^3>(\tau') d\tau' d\tau \; .
\label{four1}
\end{align}

The first and second terms in (\ref{four1}) arise in our analysis but are absent in Ref. \cite{PRL}. We have already encountered these terms in Appendix \ref{calculation1} while evaluating $<x^2>$ (see (\ref{square1})). The results we obtained there in (\ref{squarefinal}) show that for large values of $t$, they increase the 4th central moment $\mu_4$ by an amount,
\begin{equation}
(\Delta \mu_4)_1 \simeq \big(1 + 2\mu^2 f_0^2 \mbox{Re}\{z_1\} \big) \int_{0}^{t} M_{\psi}(\tau) d\tau \;- 2\mu^2 f_0^2 \mbox{Re}\{z_1 z_2\} \; .
\label{changemu41}
\end{equation}

The third term in (\ref{four1}) (call it $b(t)$) is just the convolution of $M_{\psi}(t)$ and $<x^2>(t)$ and hence, using (\ref{square3}), one can write,
\begin{align}
\nonumber b(t) \simeq& \;6 \mathcal{L}^{-1} \bigg[(1 + \mu^2 f_0^2 \mbox{Re}\{z_1\}) \frac{(\hat{M}_{\psi}(s))^2}{s} - \mu^2 f_0^2 \mbox{Re}\{z_1 z_2\} \frac{\hat{M}_{\psi}(s)}{s}\bigg] \\
=& \;6\big(1 + \mu^2 f_0^2 \mbox{Re}\{z_1\}\big) \int_{0}^{t} M_{\psi}(\tau) \ast M_{\psi}(\tau) d\tau - 6\mu^2 f_0^2 \mbox{Re}\{z_1 z_2\} \int_{0}^{t} M_{\psi}(\tau) d\tau \; .
\label{four2}
\end{align}

For subdiffusive waiting time densities, $\int_{0}^{t} M_{\psi}(\tau) d\tau \sim t^\alpha$ and $\int_{0}^{t} M_{\psi}(\tau) \ast M_{\psi}(\tau) d\tau \sim t^{2\alpha}$. Hence $b(t) \sim t^{2\alpha}$.

Application of shifting and convolution property of Laplace transforms on the last term in (\ref{four1}) (call it $c(t)$) yields, after substituting results from Appendix \ref{calculation1},
\begin{equation}
c(t) = c_0(t) + c_{add}(t)\; ,
\label{bigcal1}
\end{equation}
where $c_{add}(t)$ is the contribution of the additional term in $<x^3>$, which increases the 4th central moment by $(\Delta \mu_4)_2$, given by,
\begin{equation}
(\Delta \mu_4)_2 = c_{add}(t) \simeq 2\mu^2 f_0^2 \mbox{Re}\{z_1\} \int_{0}^{t} M_{\psi}(\tau) d\tau - 2\mu^2 f_0^2 \mbox{Re}\{z_1 z_2\} \; ,
\label{bigcal2}
\end{equation}
and, $c_0(t)$ is the contribution of rest of the terms, given by,
\begin{align}
\nonumber c_0(t) \simeq& \;6 \mu^2 f_0^2 \mbox{Re}\{z_1\} \big(1 + \mu^2 f_0^2 \mbox{Re}\{z_1\}\big) \int_{0}^{t} M_{\psi}(\tau) \ast M_{\psi}(\tau) d\tau \\
\nonumber &+ 3\mu^2 f_0^2 \big[2 \mbox{Re} \{z_1^2\} - \mu^2 f_0^2 \big(\mbox{Re}\{z_1\} \mbox{Re}\{z_1 z_2\} + \mbox{Im}\{z_1\} \mbox{Im}\{z_1 z_2\}\big)\big] \int_{0}^{t} M_{\psi}(\tau) d\tau \\
&- 6 \mu^2 f_0^2 \mbox{Re}\{z_1 z_2 (z_1 + z_2)\} + 3\mu^4 f_0^4 \mbox{Re} \{z_1 z_2 [z_3(z_4 - z_2) - z_1 z_2]\} \; .
\label{bigcal3}
\end{align}
where $z_4 = \hat{M}_{\psi}(-i4\omega)$.

We know that the 4th central moment can be written as,
\begin{equation}
\mu_4 = \;<x^4> - 4 <x^3><x> + 6 <x^2><x>^2 - 3 <x>^4 \; .
\label{mu4expand}
\end{equation}

Consequently, the additional term in $<x^3>$, namely, $<x>$ (see (\ref{cube1})), directly affects $\mu_4$ through the second term in (\ref{mu4expand}) as,
\begin{equation}
(\Delta \mu_4)_3 = -4<x>^2 \;\simeq -4\mu^2 f_0^2 (\mbox{Im}\{z_1\})^2 \; .
\label{changemu43}
\end{equation}

Combining (\ref{changemu41}), (\ref{bigcal2}) and (\ref{changemu43}), we find the total increment of $\mu_4$ due to the extra terms in our analysis,
\begin{equation}
\Delta \mu_4 \simeq \big(1 + 4\mu^2 f_0^2 \mbox{Re}\{z_1\} \big) \int_{0}^{t} M_{\psi}(\tau) d\tau \;- 4\mu^2 f_0^2 \mbox{Re}\{z_1 z_2\} -  4\mu^2 f_0^2 (\mbox{Im}\{z_1\})^2 \; .
\label{changemu4}
\end{equation}

Similarly, substituting from (\ref{meanfinal}), (\ref{squarefinal}), (\ref{cubefinal}), (\ref{four2}) and (\ref{bigcal3}) into (\ref{mu4expand}), one obtains the 4th central moment as,
\begin{equation}
\mu_4 \simeq 6\big(1 + \mu^2 f_0^2 \mbox{Re}\{z_1\}\big)^2 \int_{0}^{t} M_{\psi}(\tau) \ast M_{\psi}(\tau) d\tau + A \int_{0}^{t} M_{\psi}(\tau) d\tau + B\; .
\label{bigcal4}
\end{equation}
where $A$ and $B$ are constants related to $z_i\; (i=1,2,3,4)$. For subdiffusive waiting time densities, the first term increases as $t^{2\alpha}$ and the second as $t^\alpha$. Therefore the relative increment of kurtosis is given by,
\begin{align}
\nonumber \frac{\Delta \kappa}{\kappa} =&\; \frac{\Delta \mu_4}{\mu_4}\\
\simeq&\; \frac{\big(1 + 4\mu^2 f_0^2 \mbox{Re}\{z_1\} \big) \int_{0}^{t} M_{\psi}(\tau) d\tau + C}{6\big(1 + \mu^2 f_0^2 \mbox{Re}\{z_1\}\big)^2 \int_{0}^{t} M_{\psi}(\tau) \ast M_{\psi}(\tau) d\tau + A \int_{0}^{t} M_{\psi}(\tau) d\tau + B}\; ,
\label{relkappacalculation}
\end{align}
where $C$ denotes $-4\mu^2 f_0^2 \mbox{Re}\{z_1 z_2\} -  4\mu^2 f_0^2 (\mbox{Im}\{z_1\})^2$. It is evident from this expression that the relative increment of kurtosis decays as $t^{-\alpha}$.

\end{document}